\documentclass[11pt,a4paper]{article}
\usepackage[margin=1in]{geometry}
\usepackage{amsmath,amssymb}
\usepackage{booktabs}
\usepackage{hyperref}
\usepackage{authblk}
\usepackage[dvipsnames]{xcolor}
\usepackage{tcolorbox}
\usepackage{tabularx}
\usepackage{titlesec}
\usepackage{enumitem}
\usepackage{graphicx}
\usepackage{longtable}

\tcbuselibrary{breakable,skins}
\pdfoutput=1

\hypersetup{
  colorlinks=true,linkcolor=NavyBlue,urlcolor=NavyBlue,citecolor=NavyBlue,
  pdftitle={WSCM-Lite: A Practitioner-Ready Implementation of the Weak Signal Cultivation Model},
  pdfauthor={Maurice Codourey, Emmanuel A. Gonzalez},
  pdfsubject={Human-Computer Interaction, Risk Management}
}

\titleformat{\section}{\large\bfseries}{\thesection.\quad}{0em}{}
\titleformat{\subsection}{\normalsize\bfseries}{\thesubsection\quad}{0em}{}

\definecolor{boxbg}{HTML}{EFF6FF}
\newtcolorbox{calloutbox}[1]{
  enhanced, breakable,
  colback=boxbg, colframe=NavyBlue,
  colbacktitle=NavyBlue!15, coltitle=black,
  fonttitle=\bfseries, title={#1},
  boxrule=0.6pt, titlerule=0.3pt
}
\newtcolorbox{examplebox}[1]{
  enhanced, breakable,
  colback=boxbg, colframe=NavyBlue,
  colbacktitle=NavyBlue!15, coltitle=black,
  fonttitle=\bfseries, title={#1},
  boxrule=0.6pt, titlerule=0.3pt
}

\newcolumntype{L}[1]{>{\raggedright\arraybackslash}p{#1}}
\newcolumntype{C}[1]{>{\centering\arraybackslash}p{#1}}

\newcommand{\appendixsectionformat}{%
  \titleformat{\section}{\large\bfseries}{Appendix~\thesection.\quad}{0em}{}%
}

\title{\LARGE WSCM-Lite: A Practitioner-Ready Implementation\\of the Weak Signal Cultivation Model}

\author[1]{Maurice Codourey}
\author[2]{Emmanuel A.\ Gonzalez}
\affil[1]{\small codourey@proton.me}
\affil[2]{\small emmgon@gmail.com}
\date{2026}

\begin{document}
\maketitle

% ═══════════════════════════════════════════════════════════════
\begin{abstract}
The Weak Signal Cultivation Model (WSCM) provides a mathematically rigorous framework for tracking frontline risk signals across a two-dimensional coordinate field using 15 equations and 15 tunable parameters. While this specification is designed for eventual software implementation, its computational requirements create an adoption barrier for organizations whose available infrastructure is a spreadsheet. This paper introduces WSCM-Lite, a lookup-table implementation that approximates the full WSCM's recency and decay functions to within 0.01 per step while eliminating all exponential functions, state-dependent tracking, and free parameters. Because it removes consensus momentum and reversal amplification entirely, WSCM-Lite is deliberately more conservative than the full model---escalating and de-escalating roughly one to two sessions later---rather than reproducing its coordinate trajectories exactly. The simplification replaces continuous recency weighting with a four-row lookup table, reducing the specification to eight formulas and six hardcoded constants. A 26-session worked example using the Gas Fumes signal from the parent paper demonstrates that WSCM-Lite traverses all four field regions (Question Marks $\to$ Lit Fuses $\to$ Owls $\to$ Sleeping Cats $\to$ Question Marks) and triggers SMS escalation within two sessions of the full model. Four additional scenarios validate boundary behavior, and a sensitivity analysis confirms stability under $\pm 30\%$ gap threshold variation. An accompanying Excel simulator and supplementary materials are publicly available at \url{https://github.com/axxem-ai/wscm-lite}.
\end{abstract}

\noindent\textbf{Keywords:} weak signals, risk management, human-centered design, frontline safety, spreadsheet implementation, organizational resilience

% ═══════════════════════════════════════════════════════════════
\section{Introduction}
\label{sec:intro}
% ═══════════════════════════════════════════════════════════════

The gap between knowing that a risk exists and doing something about it is rarely technological. Warning signs existed before major industrial accidents. For example, Reason (1997), Kletz and Amyotte (2019), Vaughan (1996), and Hollnagel, Woods, and Leveson (2006) have documented this pattern extensively. The Weak Signal Cultivation Model (Codourey \& Gonzalez, 2026) was designed to close this gap by giving frontline teams a structured coordinate field for tracking subtle, early-stage risk signals over time. The model builds on the Weak Signal Farming Quadrant (Codourey, 2025), which introduced the four-region metaphor and the cultivation session concept.

The full WSCM specification, published as an arXiv preprint (arXiv:2604.01495), contains 15 equations and 15 parameters. It captures second-order dynamics. Consensus momentum amplifies updates when teams consistently agree over multiple sessions, and a reversal amplifier fast-tracks directional changes when credible new evidence contradicts the prior trajectory. These mechanisms are designed for software implementation, where computation is free and precision matters.

But having a validated framework and having frontline teams actually use it are different problems. Technology adoption research consistently identifies perceived ease of use as a primary determinant of whether tools get adopted at all (Davis, 1989). Technology adoption is further shaped by organizational and environmental contexts (DePietro, Wiarda, \& Fleischer, 1990). Norman (2013) argues that when users cannot operate a tool, the design is at fault, not the user. For organizations whose available infrastructure is a shared spreadsheet and a biweekly safety meeting, the full WSCM's exponential functions, state-dependent direction tracking, and multi-parameter momentum formulas create a barrier that has nothing to do with the model's validity.

This paper introduces WSCM-Lite, a deployment simplification that addresses this barrier. The contribution is threefold:

\begin{enumerate}[nosep]
\item A \textbf{lookup-table discretization} that replaces four exponential functions with a four-row table whose weights match the full WSCM's continuous recency and decay functions to within 0.01 per step.
\item A \textbf{complete practitioner specification} requiring eight formulas, six hardcoded constants, and zero free parameters---implementable in any spreadsheet without \texttt{EXP()} functions.
\item A \textbf{computational validation} across the 26-session Gas Fumes signal from the parent paper and four additional boundary scenarios, demonstrating that WSCM-Lite preserves the same four-region journey, SMS escalation timing within two sessions, and qualitative risk narrative.
\end{enumerate}

The key design constraint is \textbf{decision-equivalence}: WSCM-Lite must produce the same management actions as the full WSCM for the same input data. Region assignments, SMS escalation triggers, and SSI severity classifications must agree closely enough that a facilitator using either tier would make the same operational decisions. The model achieves this by discretizing at points where the full WSCM's continuous functions have minimal curvature, and by accepting slightly conservative behavior (slower escalation, later SMS triggering) as the cost of eliminating computational complexity.

Both tiers produce the same data structure: timestamped $(x, y)$ coordinate pairs with associated metadata. An organization that starts with WSCM-Lite in Excel can migrate to the full WSCM in software without any data conversion. The risk locus is the same object regardless of which tier generated it.

Supplementary materials---including a step-by-step mathematics guide, a detailed 26-session worked example, and a pre-loaded Excel simulator---are publicly available at \url{https://github.com/axxem-ai/wscm-lite}.

% ═══════════════════════════════════════════════════════════════
\section{The Case for a Simplified Tier}
\label{sec:case}
% ═══════════════════════════════════════════════════════════════

\subsection{Computational Complexity as an Adoption Barrier}

The full WSCM requires four exponential function evaluations per signal per session (two recency weights $\alpha(\tau)$, $\beta(\tau)$ and two decay terms), a running direction-agreement counter with sign-change detection, and a three-term momentum formula that combines magnitude sensitivity, persistence weighting, and committee scaling. These are straightforward for a software developer but inaccessible to a safety coordinator with an Excel workbook.

Davis's (1989) Technology Acceptance Model identifies perceived ease of use as a necessary condition for adoption, independent of the tool's actual usefulness. Norman (2013) frames this more directly: complexity that serves the designer rather than the user is a design failure.

\subsection{The Procurement Problem}

Organizations considering a new risk tracking tool face a chicken-and-egg problem: they cannot justify procuring software for a framework they have not yet validated in their context, but they cannot validate the framework without a tool to run it. Saghafian et al.\ (2021), in a review of the stages of organizational technology adoption, document this pre-adoption barrier, where the inability to demonstrate value before investment blocks adoption entirely.

A zero-cost entry point breaks this cycle. WSCM-Lite allows an organization to run the model for several months using existing spreadsheet infrastructure, accumulate enough locus data to evaluate whether the framework adds value, and then make an informed procurement decision about software that implements the full WSCM.

\subsection{The Expertise Bottleneck}

Incident reporting systems fail when they demand more from frontline staff than frontline staff can deliver (Pfeiffer et al., 2010). Near-miss management systems require observability at the level where precursors are visible (Gnoni \& Saleh, 2017). The WSCM's cultivation session is deliberately low-friction---a facilitated discussion producing a few numbers per signal---but the computational layer between those numbers and the updated locus position must be equally accessible. If the facilitator cannot explain the computation to the room, trust in the output erodes.

WSCM-Lite's lookup table is explainable in one sentence: ``Look up the weight and decay from the gap type, multiply by the committee factor, and update the position.'' No exponentials, no state tracking, no momentum terms. The facilitator can verify any computation by hand during the session itself.

% ═══════════════════════════════════════════════════════════════
\section{The Full WSCM in Brief}
\label{sec:wscm}
% ═══════════════════════════════════════════════════════════════

This section summarizes the parent model to establish what WSCM-Lite simplifies. The complete specification, including all 15 equations and parameter derivations, is available in Codourey and Gonzalez (2026).

\subsection{The Cultivation Field}

The WSCM operates on a $[0, 10] \times [0, 10]$ coordinate field with two axes: Risk Intensity ($x$) and Risk Growth Potential ($y$). The field is divided at the midpoint $(5, 5)$ into four regions (Table~\ref{tab:regions}):

\begin{table}[ht]
\centering
\small
\begin{tabularx}{0.92\textwidth}{lL{2.2cm}X}
\toprule
\textbf{Region} & \textbf{Location} & \textbf{Interpretation} \\
\midrule
Question Marks & $x<5$, $y<5$ & Emerging or unclear signals requiring observation \\
Lit Fuses & $x \geq 5$, $y<5$ & High current intensity, low growth potential \\
Sleeping Cats & $x<5$, $y \geq 5$ & Low current intensity, high escalation potential \\
Owls & $x \geq 5$, $y \geq 5$ & High on both dimensions; active management required \\
\bottomrule
\end{tabularx}
\caption{The four regions of the cultivation field.}
\label{tab:regions}
\end{table}

At each cultivation session, assessors independently score each active signal on two Numeric Rating Scales (NRS, 0--4). The scores are converted to field coordinates via $x_{\text{new}} = 2.5 \times \text{mean}(\text{NRS}_x)$ and $y_{\text{new}} = 2.5 \times \text{mean}(\text{NRS}_y)$, then blended with the signal's previous position using recency-weighted exponential smoothing. Over successive sessions, the signal traces a \textbf{risk locus}---a timestamped trajectory across the field that encodes the organizational narrative of how the risk was perceived, responded to, and resolved.

\subsection{Features Designed for Software Implementation}

The full WSCM includes six computational features that WSCM-Lite removes:

\begin{enumerate}[nosep]
\item \textbf{Consensus momentum} ($\kappa$): amplifies position updates when teams consistently agree on direction over multiple sessions.
\item \textbf{Reversal amplifier} ($\rho$): accelerates repositioning when credible new evidence contradicts the prior trajectory.
\item \textbf{Continuous exponential recency weights} ($\alpha(\tau)$, $\beta(\tau)$): compute update weights as smooth functions of the dimensionless time parameter $\tau = \Delta t / T_{\text{ref}}$.
\item \textbf{$k$-counter direction tracking}: maintains a running count of consecutive same-direction updates with sign-change detection.
\item \textbf{Independent $\lambda / \nu$ time-sensitivity parameters}: allow different recency sensitivities for the $x$ and $y$ axes.
\item \textbf{15 tunable parameters}: require calibration or expert judgment to set.
\end{enumerate}

These features serve the software tier, where computation is free and the model can track subtle second-order dynamics. They are unnecessary for a spreadsheet deployment where the priority is getting teams to start tracking signals at all.

\subsection{What WSCM-Lite Preserves}

WSCM-Lite retains the core architecture that makes the WSCM useful:

\begin{itemize}[nosep]
\item The $[0, 10] \times [0, 10]$ coordinate field with four named regions
\item The NRS elicitation protocol (0--4 scales, entry constraint $\leq 1$)
\item Recency-weighted position updates (new data matters more after longer gaps)
\item Passive $y$-decay (growth urgency fades without new evidence)
\item Committee scaling (solo reporters have less repositioning authority)
\item The decay floor ($y_{\min} = 0.50$) preventing silent signal disappearance
\item SMS escalation at $d \geq 7.07$
\item The Session Severity Index (SSI) as a reporting metric
\item The risk locus as the primary analytical object
\end{itemize}

% ═══════════════════════════════════════════════════════════════
\section{The WSCM-Lite Specification}
\label{sec:spec}
% ═══════════════════════════════════════════════════════════════

\subsection{Setup: Cadence Selection}

The facilitator selects a session cadence, which sets three gap classification boundaries (Table~\ref{tab:cadence}):

\begin{table}[ht]
\centering
\small
\begin{tabular}{lccc}
\toprule
\textbf{Cadence} & \textbf{Early ($\leq$)} & \textbf{Normal ($\leq$)} & \textbf{Missed~1 ($\leq$)} \\
\midrule
Weekly & 5 days & 10 days & 21 days \\
Biweekly & 10 days & 21 days & 42 days \\
Monthly & 22 days & 45 days & 90 days \\
\bottomrule
\end{tabular}
\caption{Gap classification boundaries by cadence. Any gap exceeding the Missed~1 threshold is classified as Missed~2+.}
\label{tab:cadence}
\end{table}

This is the only configuration decision. All other parameters are hardcoded.

\subsection{Step 1: Coordinate Elicitation}

Each of the $n$ assessors ($n \geq 1$) independently scores the signal on two NRS scales (0--4):
\begin{equation}
x_{\text{new}} = 2.5 \times \text{mean}(\text{NRS}_{x}), \qquad
y_{\text{new}} = 2.5 \times \text{mean}(\text{NRS}_{y})
\label{eq:coords}
\end{equation}
\textbf{Entry constraint:} A new signal may only enter when every individual NRS score is $\leq 1$ on both axes, placing the initial position within $[0,\; 2.5] \times [0,\; 2.5]$.

\subsection{Step 2: Gap Classification and Weight Lookup}

Count the calendar days since the previous session. Classify the gap using the cadence table and look up both the recency weight $w$ and the decay factor (Table~\ref{tab:lookup}):

\begin{table}[ht]
\centering
\small
\begin{tabular}{lcc}
\toprule
\textbf{Gap type} & $\mathbf{w}$ & \textbf{decay} \\
\midrule
Early & 0.281 & 0.957 \\
Normal & 0.475 & 0.917 \\
Missed~1 & 0.700 & 0.840 \\
Missed~2+ & 0.800 & 0.770 \\
\bottomrule
\end{tabular}
\caption{Lookup table replacing the WSCM's exponential functions. Values correspond to representative gap values $\tau = 0.5, 1.0, 2.0, 3.0$ respectively, each within 0.005 of the exact exponential weight.}
\label{tab:lookup}
\end{table}

If a session was missed, $w$ increases (new data matters more because the old position is stale) and decay decreases (growth urgency has eroded more between sessions).

\subsection{Step 3: Committee Scaling}
\begin{equation}
c(n) = \min\!\bigl(1,\;\; 0.70 + 0.06 \times n\bigr)
\label{eq:cn}
\end{equation}
A solo reporter ($n = 1$) gets $c = 0.76$; a full committee ($n \geq 5$) gets $c = 1.0$. This prevents a single assessor from driving large positional shifts.

\subsection{Step 4: Effective Weight}
\begin{equation}
w_{\text{eff}} = w \times c(n)
\label{eq:weff}
\end{equation}
The effective weight ranges from 0.214 (early gap, solo reporter) to 0.800 (missed~2+ gap, full committee).

\subsection{Step 5: Position Update}

\textbf{Risk Intensity (no passive decay):}
\begin{equation}
x' = x + w_{\text{eff}} \times (x_{\text{new}} - x)
\label{eq:xupdate}
\end{equation}

\textbf{Risk Growth Potential (with passive decay):}
\begin{equation}
y_{\text{decay}} = y \times \text{decay}
\label{eq:ydecay}
\end{equation}
\begin{equation}
y' = \max\!\bigl(0.50,\;\; y_{\text{decay}} + w_{\text{eff}} \times (y_{\text{new}} - y_{\text{decay}})\bigr)
\label{eq:yupdate}
\end{equation}

If no assessment is collected (signal reviewed but not rescored): $x' = x$ and $y' = \max(0.50,\; y \times \text{decay})$.

\begin{examplebox}{Why does $y$ decay but $x$ does not?}
Risk Intensity reflects the current state of the hazard---a corroded pipe joint remains corroded whether anyone looks at it or not. Risk Growth Potential reflects urgency and momentum---if nobody reports escalation for weeks, the organizational sense of urgency naturally fades. The decay mechanism models this asymmetry.
\end{examplebox}

\subsection{Step 6: Distance and SMS Escalation}
\begin{equation}
d = \sqrt{x'^{\,2} + y'^{\,2}}
\label{eq:distance}
\end{equation}
If $d \geq 7.07$ (the distance from the origin to the midpoint $(5,5)$: $\sqrt{50} \approx 7.07$), the signal is escalated to the formal Safety Management System (SMS). The signal remains in the WSCM register for continued tracking.

\subsection{Step 7: Session Severity Index}
\begin{equation}
S = \frac{d}{14.14} \times \ln(1 + f)
\label{eq:ssi}
\end{equation}
where $f$ is the cumulative real-world occurrence count and $14.14 = \sqrt{200}$ is the maximum possible distance. The SSI is a reporting metric; it does not drive position (interpretation bands in Table~\ref{tab:ssi}).

\begin{table}[ht]
\centering
\small
\begin{tabular}{lcl}
\toprule
\textbf{SSI Range} & \textbf{Status} & \textbf{Recommended Action} \\
\midrule
$[0.0,\,0.5)$ & Low & Routine monitoring \\
$[0.5,\,1.5)$ & Moderate & Team review recommended \\
$[1.5,\,2.5)$ & Elevated & Management notification \\
$\geq 2.5$ & Critical & Escalation imminent or SMS triggered \\
\bottomrule
\end{tabular}
\caption{SSI interpretation bands.}
\label{tab:ssi}
\end{table}

\subsection{Signal Retirement}

A signal may be retired when $d < 1.0$ and the team makes an active decision to close it. The decay floor ensures $y$ never reaches zero autonomously, so retirement always requires human judgment. The risk locus is retained permanently.

\begin{calloutbox}{Complete Parameter Reference}
\small
Six hardcoded constants: $n_{\text{ref}} = 5$ (full committee), $\alpha_{\min} = 0.70$ (solo reporter floor $\to c(1) = 0.76$), $y_{\min} = 0.50$ (decay floor), $d_{\text{SMS}} = 7.07$ (escalation threshold), $d_{\text{close}} = 1.0$ (retirement eligibility), and $d_{\max} = 14.14$ (SSI normalization). No free parameters require facilitator tuning. All recency weights and decay factors are fixed in the lookup table.
\end{calloutbox}

% ═══════════════════════════════════════════════════════════════
\section{Running a Cultivation Session with WSCM-Lite}
\label{sec:session}
% ═══════════════════════════════════════════════════════════════

\begin{figure}[ht]
\centering
\includegraphics[width=0.45\textwidth]{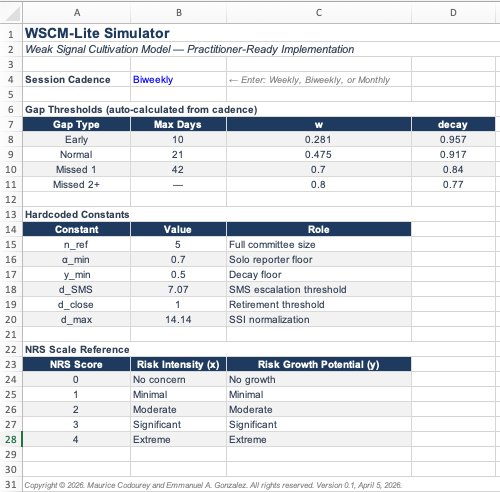}
\caption{The WSCM-Lite Config sheet showing cadence selection, gap thresholds, hardcoded constants, and NRS scale reference.}
\label{fig:config}
\end{figure}

The WSCM-Lite Excel simulator consists of four sheets: Config (Figure~\ref{fig:config}), Signal Tracker (Figure~\ref{fig:tracker}), SSI Chart, and Instructions. The simulator is available for download at \url{https://github.com/axxem-ai/wscm-lite}.

\subsection{Before the Session}

The facilitator prepares the Signal Tracker spreadsheet with the current session's day number and reviews the previous session's positions. For each active signal, the facilitator notes the current region, distance, and any pending actions from the prior session.

\subsection{During the Session}

The session follows a structured protocol for each active signal:

\begin{enumerate}[nosep]
\item \textbf{Field report.} The facilitator asks: ``What have you observed about this signal since our last session?'' Workers describe observations, near-misses, changes in frequency, or new information.
\item \textbf{NRS scoring.} Each assessor independently scores the signal on both axes (0--4). Scores are entered into the spreadsheet. The facilitator does not influence scoring.
\item \textbf{Computation.} The spreadsheet auto-computes $x_{\text{new}}$, $y_{\text{new}}$, gap classification, $w$, decay, $c(n)$, $w_{\text{eff}}$, updated position $(x', y')$, distance $d$, SMS status, and SSI.
\item \textbf{Interpretation.} The facilitator reads the result: ``The signal moved from $(3.55, 3.84)$ to $(4.46, 3.40)$. It is still in Question Marks but approaching the Lit Fuses boundary. SSI is 0.91, Moderate.''
\item \textbf{Decision.} The team determines actions: continue monitoring, escalate a maintenance request, brief management, restrict access, or other operational responses proportional to the signal's position and trajectory.
\item \textbf{New signals.} Any worker may propose a new signal. If accepted, the entry constraint is verified and the signal is placed at its initial coordinates.
\end{enumerate}

\subsection{After the Session}

\begin{figure}[ht]
\centering
\includegraphics[width=\textwidth]{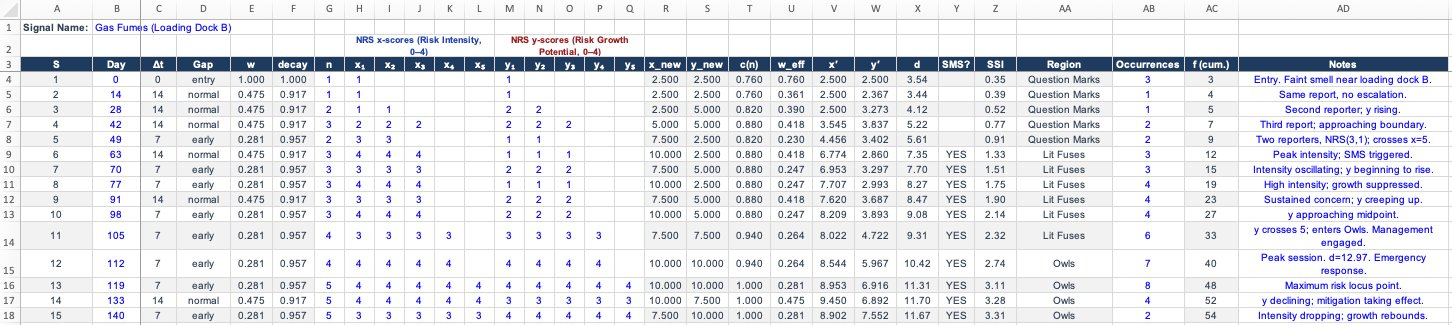}
\caption{The WSCM-Lite Signal Tracker populated with the Gas Fumes example (Sessions 1--15 shown). Blue text indicates user inputs; black text is auto-computed. The full 26-session dataset is available in the downloadable simulator.}
\label{fig:tracker}
\end{figure}

The team has produced one new row per active signal in the tracker (see Figure~\ref{fig:tracker}), an updated position and SSI for each signal, and a decision record tied to specific coordinates. The dataset is directly compatible with the full WSCM---the coordinate pairs, timestamps, NRS scores, and metadata are the same objects that a WSCM software implementation would ingest.

% ═══════════════════════════════════════════════════════════════
\section{Illustrative Application: The Risk Journey of Gas Fumes}
\label{sec:gasfumes}
% ═══════════════════════════════════════════════════════════════

The following example traces the lifecycle of a single weak signal---intermittent gas fumes reported near a loading dock---across 26 cultivation sessions spanning 252 days (36 weeks). This example uses the same NRS inputs as the parent paper's Gas Fumes trajectory (Codourey \& Gonzalez, 2026, Section~5 and Table~8). Both tiers are driven by the identical session-by-session scores and day schedule and differ only in the update model, so any divergence in the resulting coordinates isolates the effect of WSCM-Lite's lookup-table simplification. Default biweekly cadence applies.

A complete session-by-session narrative with field observations, assessor scores, step-by-step calculations, and team decisions for all 26 sessions is available in the supplementary materials at \url{https://github.com/axxem-ai/wscm-lite}. Two representative sessions are worked in full below to demonstrate the computation. Displayed intermediate values are rounded to two decimal places (round half up); the underlying spreadsheet computation carries full precision.

\subsection{Worked Example: Session 1 (Entry)}

A shift worker reports a faint, intermittent chemical smell near loading dock~B. The smell has been noticed three times over the past month before anyone brought it to a session. One assessor ($n = 1$) scores NRS$(1, 1)$.

\begin{align*}
x_{\text{new}} &= 2.5 \times 1.0 = 2.50, \qquad y_{\text{new}} = 2.5 \times 1.0 = 2.50
\end{align*}

This is the entry session. The entry constraint is satisfied (all NRS scores $\leq 1$). The signal is placed directly at the computed coordinates:

\begin{align*}
x' &= 2.50, \qquad y' = 2.50 \\
d &= \sqrt{2.50^2 + 2.50^2} = 3.54 \\
f &= 3, \qquad S = \frac{3.54}{14.14} \times \ln(1+3) = 0.250 \times 1.386 = 0.35 \;\text{(Low)}
\end{align*}

\noindent Position: $(2.50, 2.50)$ --- Question Marks. SMS: No ($d < 7.07$). Team decision: log the signal and monitor.

\subsection{Worked Example: Session 6 (SMS Triggered)}

Day~63. A gas detection survey confirms elevated methane readings in the sub-floor utility chase. Three assessors (Workers~A, B, and the safety officer) all score NRS$(4, 1)$---maximum intensity, low growth.

\begin{align*}
x_{\text{new}} &= 2.5 \times 4.0 = 10.00, \qquad y_{\text{new}} = 2.5 \times 1.0 = 2.50
\end{align*}

Gap: 14~days $\to$ Normal. $w = 0.475$, decay $= 0.917$. Committee scaling: $c(3) = \min(1, 0.70 + 0.18) = 0.88$. Effective weight: $w_{\text{eff}} = 0.475 \times 0.88 = 0.418$.

Previous position from Session~5: $(4.46, 3.40)$.

\begin{align*}
x' &= 4.46 + 0.418 \times (10.00 - 4.46) = 4.46 + 2.32 = 6.77 \\
y_{\text{decay}} &= 3.40 \times 0.917 = 3.12 \\
y' &= \max(0.50,\; 3.12 + 0.418 \times (2.50 - 3.12)) = \max(0.50,\; 2.86) = 2.86 \\
d &= \sqrt{6.77^2 + 2.86^2} = 7.35
\end{align*}

$d = 7.35 \geq 7.07$ --- \textbf{SMS escalation triggered.}

$f = 12$, $S = \frac{7.35}{14.14} \times \ln(13) = 0.520 \times 2.565 = 1.33$ (Moderate).

\noindent Position: $(6.77, 2.86)$ --- Lit Fuses. The signal has crossed $x = 5$ and the SMS threshold simultaneously. Team decision: initiate formal incident investigation, restrict dock~B access, install continuous monitoring.

\subsection{Phase 1: Question Marks (Sessions 1--5, Days 0--49)}

The signal enters at $(2.50, 2.50)$ when a single worker reports a faint, intermittent chemical smell near loading dock~B. Three prior occurrences went unlogged. Over the next four sessions, a second and third reporter join, NRS scores gradually increase, and the signal's $x$ coordinate approaches the $x = 5$ boundary. By Session~5, the position is $(4.46, 3.40)$ with SSI = 0.91 (Moderate).

\subsection{Phase 2: Lit Fuses (Sessions 6--11, Days 63--105)}

At Session~6, a gas detection survey confirms elevated methane readings in the sub-floor utility chase. All three assessors score maximum intensity (NRS$_x$ = 4). The signal jumps to $(6.77, 2.86)$, crossing $x = 5$ into Lit Fuses and triggering SMS escalation at $d = 7.35$. Over the next five sessions, a corroded gas line joint is found but repair parts are delayed. The signal oscillates in the right half of the field as intensity remains high. The $y$ coordinate creeps upward from 2.86 to 4.72 as the unrepaired source raises growth concerns. A fourth assessor (maintenance supervisor) joins at Session~11.

\subsection{Phase 3: Owls (Sessions 12--20, Days 112--182)}

At Session~12, the corroded joint partially separates during a temperature spike. Gas alarms activate. The area is evacuated. All four assessors score $(4, 4)$. The signal enters Owls at $(8.54, 5.97)$ with SSI = 2.74 (Critical). Session~13 brings a full committee of five assessors and peak distance $d = 11.31$.

Emergency repairs proceed through Sessions 14--17. The permanent repair is installed at Session~15. Both coordinates begin declining. By Session~19, intensity has dropped to $x = 5.42$ but the team maintains high growth scores because summer peak temperatures have not yet arrived. Session~20 sees a brief $y$ rebound to 7.85 when a trace detection occurs during the hottest day of the month.

\subsection{Phase 4: Sleeping Cats (Sessions 21--24, Days 189--224)}

At Session~21, $x$ drops below 5 while $y$ remains above 5---the signal enters Sleeping Cats at $(4.57, 8.13)$. The team maintains high growth scores after learning of a similar corrosion failure at another facility. A facility-wide corrosion survey is initiated, and two additional joints are flagged for preventive replacement.

By Session~24, with no detections for four consecutive weeks and preventive replacements completed, the signal drops below the SMS threshold ($d = 6.22$) for the first time since Session~6.

\subsection{Phase 5: Return to Question Marks (Sessions 25--26, Days 238--252)}

The signal crosses back below $y = 5$, returning to Question Marks. Session~26 finds the signal at $(2.71, 3.22)$ with $d = 4.21$ and SSI = 1.21 (Moderate). The team designates it a retirement candidate, noting that $d = 4.21 > 1.0$ means it does not yet meet the formal retirement threshold. The accumulated history of 58 real-world occurrences and months of elevated growth scores keeps $y$ above 3.0 even after eight weeks of clean readings. The complete 26-session trajectory is summarized in Table~\ref{tab:gasfumes}.

\begin{table}[t]
\centering
\scriptsize
\setlength{\tabcolsep}{3pt}
\begin{tabular}{ccccclcccl}
\toprule
\textbf{S} & \textbf{Day} & $\mathbf{x'}$ & $\mathbf{y'}$ & $\mathbf{n}$ & \textbf{Region} & \textbf{Occ.} & $\mathbf{f}$ & $\mathbf{S}$ & \textbf{Key event} \\
\midrule
1 & 0 & 2.50 & 2.50 & 1 & QM & 3 & 3 & 0.35 & Entry \\
2 & 14 & 2.50 & 2.37 & 1 & QM & 1 & 4 & 0.39 & Same reporter \\
3 & 28 & 2.50 & 3.27 & 2 & QM & 1 & 5 & 0.52 & Second reporter \\
4 & 42 & 3.55 & 3.84 & 3 & QM & 2 & 7 & 0.77 & Third reporter \\
5 & 49 & 4.46 & 3.40 & 2 & QM & 2 & 9 & 0.91 & Intensity spike \\
6 & 63 & 6.77 & 2.86 & 3 & LF & 3 & 12 & 1.33 & \textbf{SMS triggered} \\
7 & 70 & 6.95 & 3.30 & 3 & LF & 3 & 15 & 1.51 & Oscillating \\
8 & 77 & 7.71 & 2.99 & 3 & LF & 4 & 19 & 1.75 & Peak hot day \\
9 & 91 & 7.62 & 3.69 & 3 & LF & 4 & 23 & 1.90 & Source found \\
10 & 98 & 8.21 & 3.89 & 3 & LF & 4 & 27 & 2.14 & Repair delayed \\
11 & 105 & 8.02 & 4.72 & 4 & LF & 6 & 33 & 2.32 & Near boundary \\
12 & 112 & 8.54 & 5.97 & 4 & OW & 7 & 40 & 2.74 & \textbf{Emergency} \\
13 & 119 & 8.95 & 6.92 & 5 & OW & 8 & 48 & 3.11 & \textbf{Peak $d$} \\
14 & 133 & 9.45 & 6.89 & 5 & OW & 4 & 52 & 3.28 & Bypass holding \\
15 & 140 & 8.90 & 7.55 & 5 & OW & 2 & 54 & 3.31 & \textbf{Peak SSI} \\
16 & 147 & 8.53 & 7.30 & 4 & OW & 1 & 55 & 3.20 & Repair installed \\
17 & 154 & 7.66 & 7.11 & 3 & OW & 1 & 56 & 2.99 & Source fixed \\
18 & 161 & 6.38 & 6.98 & 3 & OW & 1 & 57 & 2.72 & Readings dropping \\
19 & 168 & 5.42 & 6.88 & 3 & OW & 0 & 57 & 2.52 & No detections \\
20 & 182 & 5.25 & 7.85 & 3 & OW & 1 & 58 & 2.72 & Summer caution \\
21 & 189 & 4.57 & 8.13 & 3 & SC & 0 & 58 & 2.69 & Enters Sl.\ Cats \\
22 & 196 & 4.06 & 7.71 & 3 & SC & 0 & 58 & 2.51 & Survey started \\
23 & 210 & 3.41 & 6.21 & 3 & SC & 0 & 58 & 2.04 & Survey done \\
24 & 224 & 3.05 & 5.42 & 2 & SC & 0 & 58 & 1.79 & \textbf{Below SMS} \\
25 & 238 & 2.84 & 4.01 & 2 & QM & 0 & 58 & 1.42 & Returns to QM \\
26 & 252 & 2.71 & 3.22 & 2 & QM & 0 & 58 & 1.21 & Retirement cand. \\
\bottomrule
\end{tabular}
\caption{Gas Fumes signal: 26-session WSCM-Lite trajectory. QM = Question Marks, LF = Lit Fuses, OW = Owls, SC = Sleeping Cats. SMS active Sessions 6--23.}
\label{tab:gasfumes}
\end{table}

\begin{figure}[ht]
\centering
\includegraphics[width=0.85\textwidth]{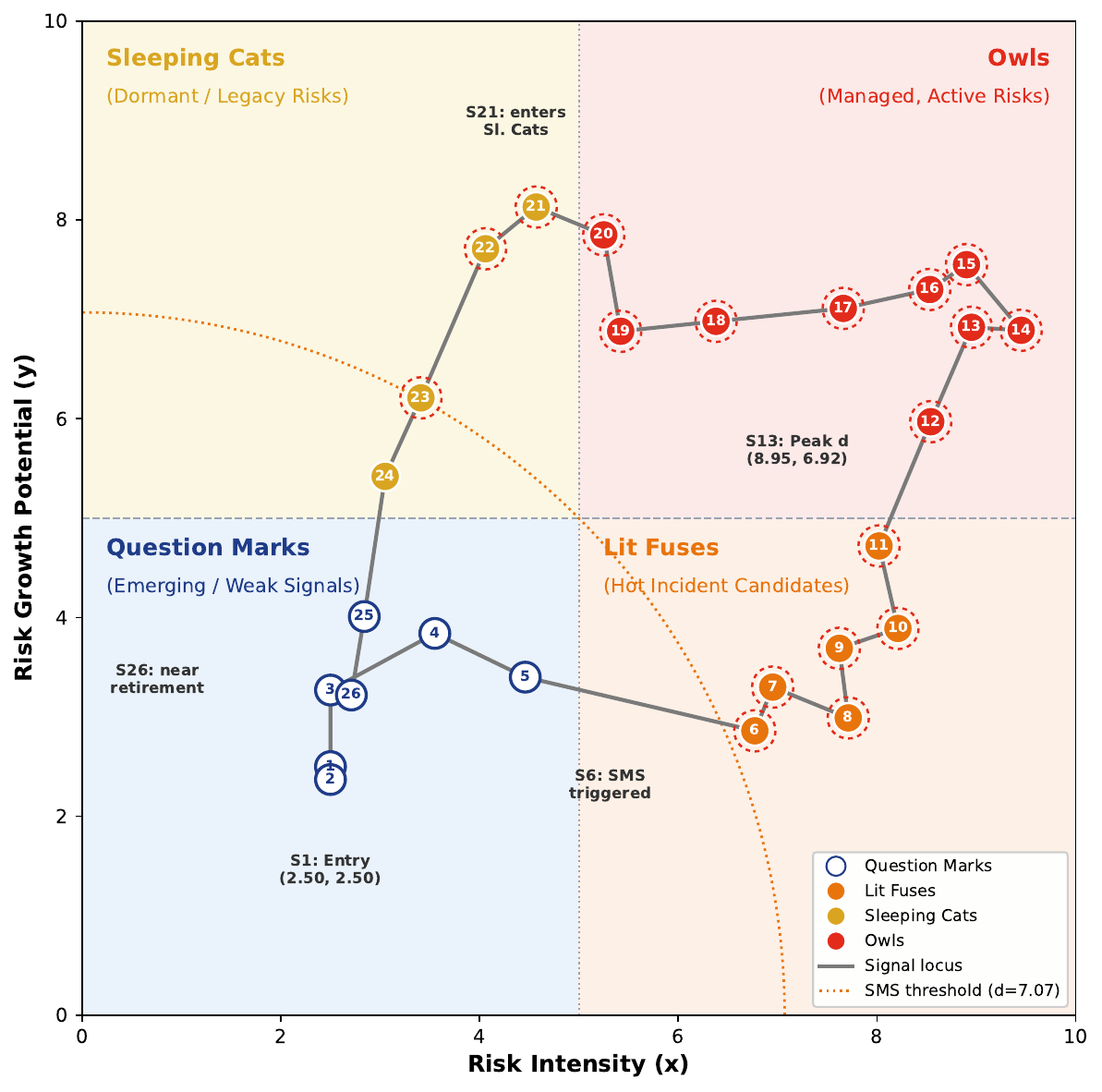}
\caption{Gas Fumes risk locus across 26 cultivation sessions. The signal traverses all four field regions: Question Marks $\to$ Lit Fuses $\to$ Owls $\to$ Sleeping Cats $\to$ Question Marks.}
\label{fig:locus}
\end{figure}

The risk locus (Figure~\ref{fig:locus}) visualizes the signal's trajectory across all four regions of the cultivation field. Each node is labeled with its session number and color-coded by region. Dashed red rings indicate sessions where SMS escalation was active ($d \geq 7.07$). The trajectory clearly shows the escalation arc (S1--S13), the crisis response phase (S13--S20), and the gradual de-escalation back toward Question Marks (S21--S26).

\begin{figure}[ht]
\centering
\includegraphics[width=0.95\textwidth]{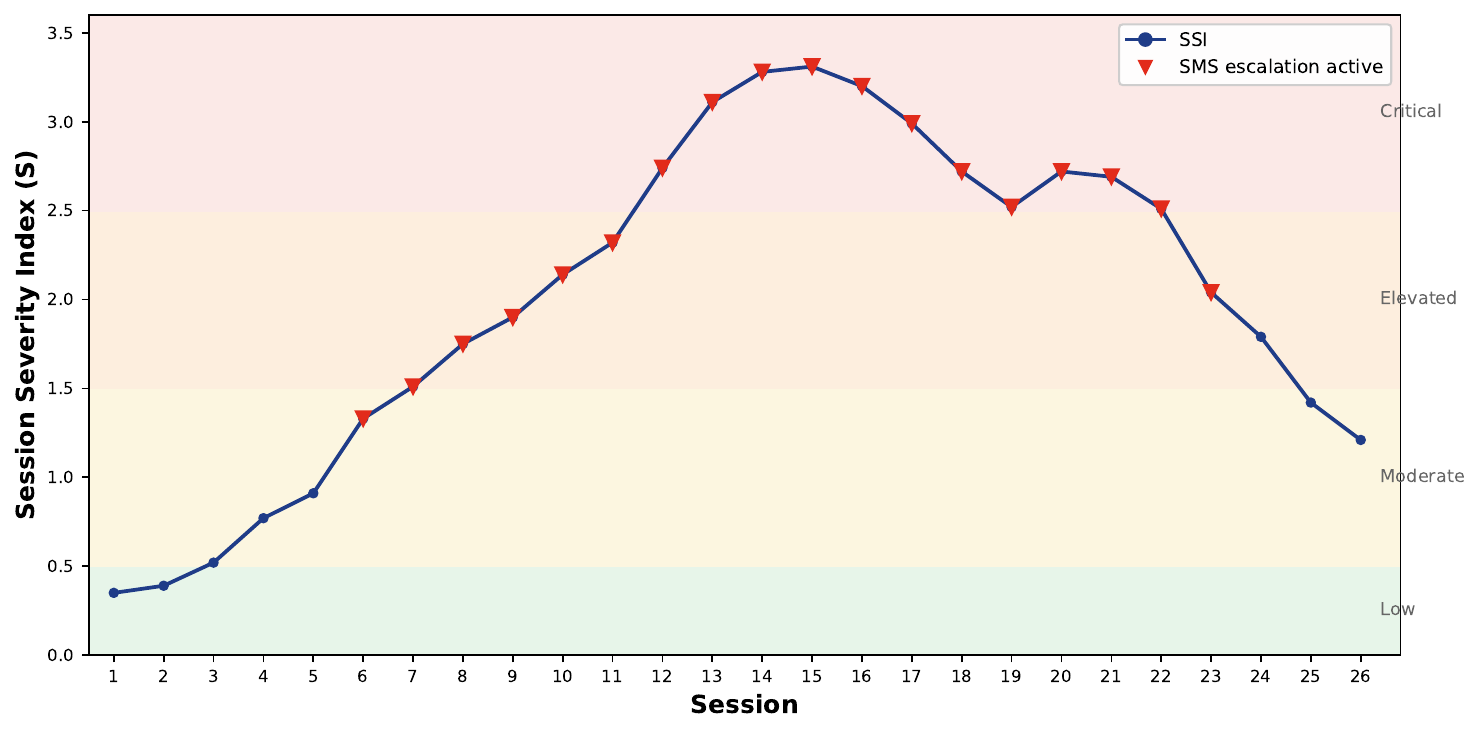}
\caption{Session Severity Index (SSI) over time for the Gas Fumes signal. Colored bands correspond to the interpretation thresholds in Table~\ref{tab:ssi}. Red triangles indicate sessions with SMS escalation active.}
\label{fig:ssi}
\end{figure}

The Session Severity Index over time (Figure~\ref{fig:ssi}) shows how the combined effect of distance and cumulative frequency produces a severity curve that peaks at Session~15 ($S = 3.31$, Critical) and remains elevated even as the signal's coordinates decline, because the accumulated occurrence count ($f = 58$) sustains the SSI through the $\ln(1+f)$ term.

% ═══════════════════════════════════════════════════════════════
\section{Validation: Alternative Scenarios}
\label{sec:scenarios}
% ═══════════════════════════════════════════════════════════════

To validate WSCM-Lite's behavior beyond the Gas Fumes trajectory, four additional scenarios were simulated. Each tests a distinct boundary condition.

\subsection{Scenario A: The Quiet Signal}

A signal enters at $(2.50, 2.50)$ and is scored $(1, 1)$ by a three-member committee at every biweekly session for 20 sessions. With $w = 0.475$ (Normal gap) and consistent low scores, the $x$ coordinate holds at 2.50 while $y$ settles to a fixed point where passive decay balances the low-but-nonzero input. By Session~20, the signal converges to approximately $(2.50, 2.24)$---still in Question Marks, never approaching any boundary. Because the steady input keeps $y$ well above the decay floor ($y_{\min} = 0.50$), the floor is never engaged. This confirms that a genuinely low-risk signal does not drift into escalation through model artifacts.

\subsection{Scenario B: The Oscillating Signal}

A signal alternates between moderate scores $(2, 2)$ and low scores $(1, 1)$ every other session. Without the full WSCM's momentum mechanism, WSCM-Lite treats each session independently. The signal oscillates in the lower-center of the field, its intensity peaking near $x = 4.1$ and never reaching the Lit Fuses boundary at $x = 5$. This is the expected cost of removing consensus momentum: oscillating signals are treated as uncertain rather than escalating.

\subsection{Scenario C: Rapid Reversal}

A signal escalates quickly (NRS$(4, 4)$ for three sessions) then reverses sharply (NRS$(1, 1)$ for three sessions). The full WSCM's reversal amplifier would accelerate the de-escalation. WSCM-Lite de-escalates at the standard rate, resulting in a 1--2 session delay in crossing back below the SMS threshold. The qualitative outcome is identical: the signal returns to Question Marks.

\subsection{Scenario D: Solo Reporter}

A single assessor ($n = 1$) tracks a signal for 15 sessions with gradually increasing scores. The committee scaling factor $c(1) = 0.76$ limits $w_{\text{eff}}$ to at most $0.76 \times 0.800 = 0.608$. The signal moves somewhat more slowly than it would with a full committee. Under a gradual ramp this difference is small---typically zero to one session in reaching Lit Fuses---because both committee sizes cross $x = 5$ once the inputs grow large enough; the committee-scaling conservatism is most pronounced under abrupt escalation rather than gradual drift. This is the intended behavior: a solo reporter should not drive the same organizational response as a committee consensus. However, this scenario also reveals a limitation---if only one person is reporting and their concern is legitimate, the model structurally underweights it.

\subsection{Sensitivity Analysis: Gap Threshold Variation}

The gap classification boundaries were varied by $\pm 30\%$ across all three cadences. For example, the biweekly Early threshold was tested at 7 days (--30\%) and 13 days (+30\%). All variations produced maximum coordinate drift below 1.0 field units from the baseline trajectory, confirming that the lookup table discretization is stable under reasonable threshold calibration differences across organizations.

% ═══════════════════════════════════════════════════════════════
\section{Discussion}
\label{sec:discussion}
% ═══════════════════════════════════════════════════════════════

\subsection{What WSCM-Lite Gains and Loses}

The primary gain is immediate deployability. An organization can download the Excel simulator from the supplementary repository (\url{https://github.com/axxem-ai/wscm-lite}), set their cadence, and begin tracking signals in the same meeting. No software procurement, no parameter tuning, no computational expertise required.

The primary loss is responsiveness to sustained consensus and credible reversals. The full WSCM's momentum mechanism rewards teams that consistently agree over multiple sessions by amplifying updates, creating a ``flywheel effect'' that accelerates signals toward their true position. WSCM-Lite treats every session independently, so convergence is slower. Similarly, the reversal amplifier in the full WSCM allows signals to change direction quickly when new evidence is compelling. WSCM-Lite reverses at the standard rate, resulting in 1--2 session delays.

We initially considered keeping the momentum mechanism and simplifying only the recency weights. The resulting model was simpler than the full WSCM but still required state tracking across rows---the $k$-counter that records consecutive same-direction updates. That meant the facilitator could not verify the computation by hand, which defeated the purpose. So we removed momentum entirely and accepted the slower convergence as the cost of a model that fits in a lookup table.

In practice, this means WSCM-Lite is slightly more conservative than the full WSCM: it escalates later and de-escalates later. For a safety application, conservatism in de-escalation is arguably a feature---a signal that stays elevated for one extra session after conditions improve costs little, while a signal that de-escalates too quickly could lead to premature relaxation of controls.

\subsection{The Migration Path}

Both WSCM tiers produce identical data structures. An organization that has tracked signals in WSCM-Lite for six months has accumulated a dataset of timestamped $(x, y)$ pairs with NRS scores, assessor counts, and occurrence frequencies. This dataset can be ingested directly by a WSCM software implementation---no format conversion, no re-entry, no data loss.

This creates a natural procurement argument: ``We have been tracking 12 signals across 15 sessions. Here is the data. The full WSCM software would add momentum-based responsiveness and automated visualization. Here is what it would cost.'' The WSCM-Lite period functions as a proof-of-concept that de-risks the software investment.

\subsection{Limitations}

\textbf{Scenario D failure mode.} When a single reporter ($n = 1$) tracks a signal over many sessions, the committee scaling factor $c(1) = 0.76$ systematically underweights their input. If the reporter is the only person with visibility into the hazard (e.g., a night-shift worker), the model structurally slows the signal's movement toward escalation. This is a known trade-off: the committee scaling protects against idiosyncratic overreaction at the cost of delaying legitimate single-source warnings.

\textbf{No adaptive cadence.} WSCM-Lite uses fixed gap boundaries. The full WSCM allows organizations to adjust $T_{\text{ref}}$ dynamically---for example, switching from biweekly to weekly sessions during a crisis. WSCM-Lite supports this by selecting a different cadence column, but the switch is manual and does not automatically recalibrate historical gap classifications.

\textbf{Lookup table granularity.} The four-bucket discretization ($\tau \approx 0.5, 1.0, 2.0, 3.0$) is optimized for regular-cadence sessions. Organizations with highly irregular meeting schedules may experience gaps that fall in the transition zone between buckets, where the rounding error is largest. The sensitivity analysis confirms this error stays below 1.0 field units even at $\pm 30\%$ threshold variation.

% ═══════════════════════════════════════════════════════════════
\section{Conclusion}
\label{sec:conclusion}
% ═══════════════════════════════════════════════════════════════

WSCM-Lite demonstrates that the core contribution of the Weak Signal Cultivation Model---giving frontline teams a structured, two-dimensional coordinate field for tracking risk signals over time---does not require computational sophistication. The lookup-table implementation matches the full model's recency and decay functions to within 0.01 per step, preserves the same four-region journey across all tested scenarios, and triggers SMS escalation within two sessions of the full model.

The eight formulas and six constants presented here can be implemented in any spreadsheet. No exponential functions. No tunable parameters. No software beyond what every organization already has.

The value of the WSCM has always been in what it makes visible: the risk locus---a trajectory that tells the organizational story of how a signal was first noticed, how it escalated, how the organization responded, and whether the response worked. That value is zero if the tool sits unused because the math is too complex for the teams who need it most. WSCM-Lite closes this gap.

The Excel simulator, step-by-step mathematics guide, and complete 26-session worked example are available at \url{https://github.com/axxem-ai/wscm-lite}.

% ═══════════════════════════════════════════════════════════════
\vspace{18pt}
\noindent{\large\bfseries Author Contributions and AI Use Disclosure}
\vspace{6pt}

\textbf{Author contributions.} Maurice Codourey conceived the Weak Signal Cultivation Model, developed the theoretical framework, designed the coordinate field architecture and quadrant taxonomy, conducted the pilot applications, and is responsible for all conceptual and empirical content of this paper. Emmanuel A.\ Gonzalez contributed to the mathematical formalization of the coordinate derivation model, the movement scoring system, and the scientific refinement of the computational framework.

\textbf{AI use disclosure.} All ideas, theoretical contributions, model design, and research findings in this paper originate from the human authors. Claude (Opus 4.6, Anthropic) was used to draft and refine academic prose from the authors' outlines and instructions, generate Python code for figure production, recompute the illustrative simulation, and produce LaTeX typesetting for tables and document formatting. For this revised version (v2), Claude (Fable~5, Anthropic) was used to diagnose internal-consistency, citation, and numerical errors across the manuscript, and Claude (Opus~4.8, Anthropic) was used to verify those findings against source, recompute the worked examples and validation scenarios, and regenerate the figures. Grammarly was used for grammar and style review. The authors reviewed, revised, and approved all output as the final step before submission. No AI tool contributed to the conceptual design, theoretical framework, or analytical conclusions of this work. The authors take full responsibility for the accuracy, integrity, and originality of all content.

% ═══════════════════════════════════════════════════════════════
\section*{References}
\vspace{4pt}

\begin{sloppypar}
\begin{list}{}{\setlength{\leftmargin}{1.5em}\setlength{\itemindent}{-1.5em}\setlength{\itemsep}{4pt}\setlength{\parsep}{0pt}}

\item Codourey, M. (2025). Weak signal farming quadrant---A human-centric approach for frontline risk detection and resilience. ResearchGate. \url{https://www.researchgate.net/publication/395334608}

\item Codourey, M., \& Gonzalez, E.\,A. (2026). The weak signal cultivation model: A human-centric framework for frontline risk detection, signal tracking, and proactive organizational resilience. \textit{arXiv preprint} arXiv:2604.01495. \url{https://arxiv.org/abs/2604.01495}

\item Davis, F.\,D. (1989). Perceived usefulness, perceived ease of use, and user acceptance of information technology. \textit{MIS Quarterly}, 13(3), 319--340.

\item DePietro, R., Wiarda, E., \& Fleischer, M. (1990). The context for change: Organization, technology and environment. In L.\,G.\ Tornatzky \& M.\ Fleischer (Eds.), \textit{The Processes of Technological Innovation} (pp.\ 151--175). Lexington Books.

\item Gnoni, M.\,G., \& Saleh, J.\,H. (2017). Near-miss management systems and observability-in-depth: Handling safety incidents and accident precursors in light of safety principles. \textit{Safety Science}, 91, 154--167.

\item Gonzalez, E.\,A., Presto, R.\,S., Remacha, A.\,C., \& Santos, A.\,N. (2015). A model describing hazard identification effectiveness of workers in the construction and maintenance industry. In \textit{Proceedings of the 8th IEEE GCC Conference and Exhibition}. IEEE. \url{https://doi.org/10.1109/IEEEGCC.2015.7060044}

\item Hollnagel, E., Woods, D.\,D., \& Leveson, N. (2006). \textit{Resilience Engineering: Concepts and Precepts}. Ashgate.

\item Kletz, T., \& Amyotte, P. (2019). \textit{What Went Wrong? Case Histories of Process Plant Disasters and How They Could Have Been Avoided} (6th ed.). Elsevier.

\item Norman, D.\,A. (2013). \textit{The Design of Everyday Things} (Revised and expanded ed.). Basic Books.

\item Pfeiffer, Y., Manser, T., \& Wehner, T. (2010). Conceptualising barriers to incident reporting: A psychological framework. \textit{Quality and Safety in Health Care}, 19(6), e60.

\item Reason, J. (1997). \textit{Managing the Risks of Organizational Accidents}. Ashgate.

\item Roberts, S.\,W. (1959). Control chart tests based on geometric moving averages. \textit{Technometrics}, 1(3), 239--250.

\item Saghafian, M., Laumann, K., \& Skogstad, M.\,R. (2021). Stagewise overview of issues influencing organizational technology adoption and use. \textit{Frontiers in Psychology}, 12, 630145.

\item Vaughan, D. (1996). \textit{The Challenger Launch Decision: Risky Technology, Culture, and Deviance at NASA}. University of Chicago Press.

\end{list}
\end{sloppypar}

% ═══════════════════════════════════════════════════════════════
\appendix
\appendixsectionformat

\section{Lookup Table Derivation}
\label{app:lookup}

The four lookup table values are the WSCM's continuous exponential functions evaluated at a representative $\tau$ value for each gap bucket. The underlying exponential moving average approach follows Roberts (1959).

The recency weight in the full WSCM is:
\[
w(\tau) = \alpha_{\text{base}} \times (1 - e^{-\lambda \tau})
\]
with $\alpha_{\text{base}} = 0.90$ and $\lambda = 0.75$. Evaluating at four representative $\tau$ values yields the recency weights in Table~\ref{tab:wderiv}.

\begin{table}[ht]
\centering
\small
\begin{tabular}{lcccc}
\toprule
\textbf{Gap type} & $\boldsymbol{\tau}$ & \textbf{Exact $w(\tau)$} & \textbf{Rounded} & \textbf{Error} \\
\midrule
Early & 0.5 & 0.28144 & 0.281 & $< 0.001$ \\
Normal & 1.0 & 0.47487 & 0.475 & $< 0.001$ \\
Missed~1 & 2.0 & 0.69918 & 0.700 & $0.001$ \\
Missed~2+ & 3.0 & 0.80514 & 0.800 & $0.005$ \\
\bottomrule
\end{tabular}
\caption{Recency weight derivation from the WSCM exponential function.}
\label{tab:wderiv}
\end{table}

The passive $y$-decay factor is $e^{-\mu \tau}$ with $\mu = 0.087$ (Table~\ref{tab:decayderiv}):

\begin{table}[ht]
\centering
\small
\begin{tabular}{lcccc}
\toprule
\textbf{Gap type} & $\boldsymbol{\tau}$ & \textbf{Exact decay} & \textbf{Rounded} & \textbf{Error} \\
\midrule
Early & 0.5 & 0.95744 & 0.957 & $< 0.001$ \\
Normal & 1.0 & 0.91669 & 0.917 & $< 0.001$ \\
Missed~1 & 2.0 & 0.84032 & 0.840 & $< 0.001$ \\
Missed~2+ & 3.0 & 0.77030 & 0.770 & $< 0.001$ \\
\bottomrule
\end{tabular}
\caption{Decay factor derivation from the WSCM exponential function.}
\label{tab:decayderiv}
\end{table}

The maximum single-step rounding error is $0.005 \times 10 = 0.05$ field units (worst case: Missed~2+ gap, full committee, maximum coordinate difference). The error retention factor is $1 - \min(w_{\text{eff}}) = 1 - 0.214 = 0.786$, giving a geometric series bound of $0.05 / (1 - 0.786) \approx 0.23$ field units. In the 26-session Gas Fumes simulation, the observed maximum accumulated drift was 0.009 field units---well below the theoretical bound.

% ═══════════════════════════════════════════════════════════════

\vspace{14pt}
\noindent\rule{\columnwidth}{0.4pt}
\vspace{4pt}
\noindent{\footnotesize\textcopyright\ Codourey \& Gonzalez, 2026. Licensed under CC BY-NC 4.0.}

\end{document}